\documentclass[article,preprint,12pt,unsrt]{elsarticle}
\usepackage{a4wide}
\usepackage{graphicx}
\usepackage{amsfonts}
\usepackage{amsmath}
\usepackage{booktabs}
\usepackage{epsfig}
\usepackage{stmaryrd}
\usepackage{lscape}
\usepackage{hyperref}
\hypersetup{citecolor=black,linkcolor= black}

\begin{document}
\title{Fractal approach towards power-law coherency to measure cross-correlations between time series}
\author{Ladislav Kristoufek}
\ead{kristouf@utia.cas.cz}
\address{Institute of Information Theory and Automation, Czech Academy of Sciences, Pod Vodarenskou Vezi 4, CZ-182 08, Prague 8, Czech Republic
%\\
%Institute of Economic Studies, Faculty of Social Sciences, Charles University, Opletalova 26, 110 00, Prague 1, Czech Republic
}

\begin{abstract}
We focus on power-law coherency as an alternative approach towards studying power-law cross-correlations between simultaneously recorded time series. To be able to study empirical data, we introduce three estimators of the power-law coherency parameter $H_{\rho}$ based on popular techniques usually utilized for studying power-law cross-correlations -- detrended cross-correlation analysis (DCCA), detrending moving-average cross-correlation analysis (DMCA) and height cross-correlation analysis (HXA). In the finite sample properties study, we focus on the bias, variance and mean squared error of the estimators. We find that the DMCA-based method is the safest choice among the three. The HXA method is reasonable for long time series with at least $10^4$ observations, which can be easily attainable in some disciplines but problematic in others. The DCCA-based method does not provide favorable properties which even deteriorate with an increasing time series length. The paper opens a new venue towards studying cross-correlations between time series.
\end{abstract}


\journal{Communications in Nonlinear Science and Numerical Simulation}

\maketitle
%
%
%\textit{PACS codes: }\\

\newpage

\section{Introduction}

Analyzing fractal properties of time series has been a significant contribution of physics to a broad range of other disciplines \cite{Zebende2009,Zhao2011,Stan2013,Jun2012,Shadkhoo2009,Marinho2013,Xue2012, Ursilean2009,Hajian2010,Vassoler2012,Kang2013,He2011,Wang2013b,Podobnik2009a,Shao2012}. In time series, fractal properties translate directly into specific correlation structures. Hurst exponent $H$, as a characteristic parameter of fractal series, provides an insight into asymptotic scaling of the auto-correlation function, specifically its power-law decay. In the case of stationary series, we have $0 \le H<1$ with a separating point of $H=0.5$ characteristic for uncorrelated series. Persistent or long-range correlated series with $H>0.5$ follow local trends but still remain mean reverting and stationary, while anti-persistent series with $H<0.5$ are distinctive by their excessive switching (with respect to uncorrelated processes) \cite{Beran1994}.

Recently, the methodological framework has been generalized into the bivariate setting so that not only long-range correlations but also long-range cross-correlations can be studied using methods developed in physics \cite{Meneveau1990,Wang2012,Xie2015,Oswiecimka2014,Kwapien2015,Qian2015,Gu2010}. The most popular ones have been the detrended fluctuation analysis (DFA) \cite{Peng1993,Peng1994} and the detrended cross-correlation analysis (DCCA or DXA) \cite{Podobnik2008,Zhou2008,Jiang2011} as its bivariate generalization. The development of DCCA has motivated others in introducing alternative methods such as the detrending moving-average cross-correlation analysis (DMCA) \cite{Arianos2009,He2011a} and the height cross-correlation analysis (HXA) \cite{Kristoufek2011}. These estimators of the bivariate Hurst exponent $H_{xy}$ provide an additional detail about power-law scaling of the cross-correlation function between series. Interpretation of $H_{xy}$ is usually approached through its comparison to Hurst exponents of the separate series, i.e. $H_{x}$ and $H_{y}$ \cite{Kristoufek2011,Sela2012,Kristoufek2015}.

As the next step in utilizing the fractal methods, the scale-specific correlation coefficients based on DCCA and DMCA have been proposed \cite{Zebende2011,Kristoufek2014a,Kristoufek2014b} as well as estimators of regression parameters for specific scales \cite{Kristoufek2015c,Kristoufek2016}. An important topic, which has been only slightly touched by Podobnik \textit{et al.} \cite{Podobnik2011} but not further developed, is the scaling of scale-specific correlations. However, the link between correlations scaling and bivariate Hurst exponent provides an important insight into the dynamics between time series.

Here we focus on this link in detail, building on the notion of squared spectrum coherency and translating it into the time domain, which is more frequent in the topical literature. We propose three estimators of the power-law coherency $H_{\rho}$ based on popular time domain power-law cross-correlations methods -- detrended cross-correlation analysis, detrending moving-average cross-correlation analysis and height cross-correlation analysis -- and analyze their finite sample properties. This approach presents a new way how to look at and analyze dependence between simultaneously recorded series with applications across various disciplines.

\section{Power-law coherency}

We start with a definition of the squared spectrum coherency. For two processes $\{x_t\}$ and $\{y_t\}$ with existing power spectra $f_{x}(\omega)$ and $f_{y}(\omega)$, and cross-power spectrum $f_{xy}(\omega)$ at frequency $0 < \omega \le \pi$, the squared spectrum coherency is defined as 
\begin{equation}
K_{xy}^2(\omega)=\frac{|f_{xy}(\omega)|^2}{f_{x}(\omega)f_{y}(\omega)}.
\end{equation}
If the two processes are power-law correlated so that $f_{x}(\omega) \propto \omega^{1-2H_x}$ and $f_{y}(\omega) \propto \omega^{1-2H_y}$ close to the origin ($\omega \rightarrow 0+$) with respective Hurst exponents $H_x$ and $H_y$, and in addition, the processes are power-law cross-correlated so that $|f_{xy}(\omega)| \propto \omega^{1-2H_{xy}}$ close to the origin with bivariate Hurst exponent $H_{xy}$, we have
\begin{equation}
\label{eq:coherence}
K_{xy}^2(\omega)\propto \frac{\omega^{2(1-2H_{xy})}}{\omega^{1-2H_x}\omega^{1-2H_y}}=\omega^{-4(H_{xy}-\frac{H_x+H_y}{2})} \equiv \omega ^{-4H_{\rho}}.
\end{equation}
as frequency $\omega$ approaches zero. We define the power-law coherency through parameter $H_{\rho}$ as $H_{\rho}=H_{xy}-\frac{H_x+H_y}{2}$ to respect previous discussions about the relationship between the bivariate Hurst exponent $H_{xy}$ and an average of the separate Hurst exponents \cite{Podobnik2008a,Sela2009,Sela2012,Kristoufek2013,Kristoufek2015}. Note that the squared spectrum coherency $K_{xy}^2(\omega)$ is restricted in the same way as the squared correlation, i.e. $0\le K_{xy}^2(\omega) \le 1$ for all frequencies $\omega$ \cite{Wei2006}. This yields only two possible settings for the exponent -- either $H_{\rho}=0$ or $H_{\rho}<0$ \cite{Kristoufek2015}. The former implies that the squared coherency goes to a constant and $H_{xy}=\frac{H_x+H_y}{2}$. And the latter implies that as $\omega$ approaches zero so does the squared coherency but here specifically in the power-law manner and it is thus referred to as the power-law coherency, or anti-cointegration \cite{Sela2012}, and it has $H_{xy}<\frac{H_x+H_y}{2}$. In words, such power-law coherent processes might be correlated in the short term (at high frequencies or low scales) but they are uncorrelated in the long term perspective (at low frequencies or high scales).

The squared spectrum coherency for frequency $\omega$ can be easily translated from the frequency domain to the time domain as a squared correlation for scale $s=\frac{\pi T}{\omega_j}$ parallel to frequency $\omega_j=\frac{2\pi j}{T}$ with $j=1,2,\ldots,\frac{T}{2}$ as
\begin{equation}
\label{eq_PLC_time}
\rho_{xy}^2(s)=\frac{|\sigma_{xy}(s)|^2}{\sigma^2_{x}(s)\sigma^2_{y}(s)},
\end{equation}
where $\rho_{xy}^2(s)$ is a squared correlation between $\{x_t\}$ and $\{y_t\}$ for scale $s$, and $\sigma_{xy}(s)$, $\sigma^2_{x}(s)$ and $\sigma^2_{y}(s)$ represent the scale-specific covariance and variances, respectively. The notion of power-law coherency then translates perfectly with the only difference that it occurs at high scales as a parallel to low frequencies. Specifically, we have $\sigma_{x}^2(s) \propto s^{2H_x}$ and $\sigma_{y}^2(s) \propto s^{2H_y}$ for power-law correlated processes $\{x_t\}$ and $\{y_t\}$ \cite{Beran1994,Samorodnitsky2006}, and if the processes are power-law cross-correlated, we additionally have $\sigma_{xy}(s) \propto s^{2H_{xy}}$ for $s \rightarrow +\infty$ \cite{Kristoufek2013a}. Substituting into Eq. \ref{eq_PLC_time}, we obtain
\begin{equation}
\label{eq_PLC}
\rho_{xy}^2(s) \propto \frac{s^{4H_{xy}}}{s^{2H_x}s^{2H_y}} = s^{4(H_{xy}-\frac{H_x-H_y}{2})} \equiv s^{4H_{\rho}},
\end{equation}
i.e. we have the same scaling exponent for both time ($s \rightarrow +\infty$) and frequency ($\omega \rightarrow 0+$) domain power-law coherency.

%Here we propose three estimators of the power-law coherency $H_{\rho}$ based on popular time domain power-law cross-correlations methods -- detrended cross-correlation analysis, detrending moving-average cross-correlation analysis and height cross-correlation analysis. We analyze finite sample properties of the estimators and we provide some further discussion to the topic.

\section{Estimators} %%% summarize shortly DCCA, DMCA and HXA

In this section, we recall the essentials of the bivariate Hurst exponent estimators of interest -- detrended cross-correlation analysis (DCCA), detrending moving-average cross-correlation analysis (DMCA), and height cross-correlation analysis (HXA) -- and we introduce procedures to estimate the power-law coherency parameter $H_{\rho}$ based on these.

\subsection{Bivariate Hurst exponent estimators}

In the DCCA procedure \cite{Podobnik2008}, let us consider two time series $\{x_t\}$ and $\{y_t\}$ with $t=1,\ldots,T$. Their respective profiles $\{X_t\}$ and $\{Y_t\}$, defined as $X_t=\sum_{i=1}^t{(x_i-\bar{x})}$ and $Y_t=\sum_{i=1}^t{(y_i-\bar{y})}$, for $t=1,\ldots,T$, are divided into overlapping boxes of length $s$ (scale) so that $T-s+1$ boxes are constructed. In each box between $j$ and $j+s-1$, a linear time trend is fitted so that we get $\widehat{X_{k,j}}$ and $\widehat{Y_{k,j}}$ for $j\le k \le j+s-1$. The covariance between deviations from the time trends in each box is defined as
\begin{equation}
f_{DCCA}^2(s,j)=\frac{\sum_{k=j}^{j+s-1}{(X_k-\widehat{X_{k,j}})(Y_k-\widehat{Y_{k,j}})}}{s-1}.
\label{eq:DCCA1}
\end{equation}
The covariances are finally averaged over the blocks of the same scale $s$ and the detrended covariance is obtained as
\begin{equation}
F_{DCCA}^2(s)=\frac{\sum_{j=1}^{T-s+1}{f_{DCCA}^2(s,j)}}{T-s}.
\label{eq:DCCA2}
\end{equation}
For long-range cross-correlated processes, the covariance scales as
\begin{equation}
F_{DCCA}^2(s)\propto s^{2H_{xy}}.
\label{eq:DCCA3}
\end{equation}
The power-law scaling is valid asymptotically, i.e. for high scales $s$. If we set $\{x_t\}=\{y_t\}$, the standard detrended fluctuation analysis (DFA) \cite{Peng1993,Peng1994} is obtained. The above-described procedure presents only one of possible settings as various detrending approaches can be utilized as well as non-overlapping boxes. In the simulations we present later, we stick to the linear detrending and we use non-overlapping boxes with a step of 10 due to computational feasibility.

The DMCA procedure \cite{Arianos2009,He2011a} is similar to DCCA but it differs in two important aspects. First, it is not based on the box-splitting procedure. And second, it assumes a power-law scaling of covariances with an increasing moving average window size $\kappa$. The moving average procedure can take various forms (centered, backward, forward, weighted or unweighted). We stick to the centered one as suggested in the literature \cite{Carbone2003}. Specifically, for time series $\{x_t\}$ and $\{y_t\}$ and their respective profiles $\{X_t\}$ and $\{Y_t\}$, the detrended covariance $F_{DMCA}^2(\kappa)$ is defined as
\begin{equation}
\label{eq:CC-DMA}
F_{DMCA}^2(\kappa)=\frac{1}{T-\kappa+1}\sum_{i=\lfloor\kappa/2\rfloor+1}^{T-\lfloor\kappa/2\rfloor}{\Big(X_i-\widetilde{X_i(\kappa)}\Big)\Big(Y_i-\widetilde{Y_i(\kappa)}\Big)},
\end{equation}
where $\widetilde{X_i(\kappa)}$ and $\widetilde{Y_i(\kappa)}$ are respective non-weighted centered moving averages at time point $i$ with a moving average window of length (scale) $\kappa=1,3,5,\ldots,\kappa_{max}$. Due to the centered moving average procedure, scale $\kappa_{max}$ needs to be an odd integer. For the power-law cross-correlated processes $\{x_t\}$ and $\{y_t\}$, the detrended covariance scales as
\begin{equation}
\label{eq:CC-DMA_scaling}
F_{DMCA}^2(\kappa)\propto \kappa^{2H_{xy}}.
\end{equation}
The power-law scaling is again valid for high scales represented by the moving average window $\kappa$. If we set $\{x_t\}=\{y_t\}$, the standard detrending moving average (DMA) \cite{Vandewalle1998,Alessio2002} is obtained. 

The HXA method \cite{Kristoufek2011} is based on scaling of the height-height variance function redefined for two simultaneously recorded series. Let us consider two profiles $\{X_t\}$ and $\{Y_t\}$ with time resolution $\nu$ and $t=\nu,2\nu,...,\nu\lfloor\frac{T}{\nu}\rfloor$, where $\lfloor \rfloor$ is a lower integer operator. We denote $T^{\ast}=\nu\lfloor\frac{T}{\nu}\rfloor$, which varies with $\nu$, and we label the $\tau$-lag difference as $\Delta_{\tau}X_t \equiv X_{t+\tau}-X_t$ and $\Delta_{\tau}(X_tY_t) \equiv \Delta_{\tau}X_t\Delta_{\tau}Y_t$. Height-height covariance function is defined as
\begin{equation}
\label{eq:MFHXAeq1}
K_{xy}(\tau)=\frac{\nu}{T^{\ast}}\sum_{t=1}^{T^{\ast}/\nu}\Delta_{\tau}(X_tY_t)
\end{equation}
where time interval $\tau$ generally ranges between $\nu=\tau_{min},\ldots,\tau_{max}$. Scaling relationship between $K_{xy,q}(\tau)$ and the generalized bivariate Hurst exponent $H_{xy}(q)$ is obtained as
\begin{equation}
\label{eq:MFHXAeq2}
K_{xy}(\tau) \propto \tau^{2H_{xy}}.
\end{equation}
Contrary to DCCA and DMCA, the power-law scaling is valid for low levels of differencing parameter $\tau$ \cite{DiMatteo2003,DiMatteo2005,DiMatteo2007}. HXA reduces to the height-height correlation analysis (HHCA) \cite{Barabasi1991a,Barabasi1991b} for $\{X_t\}=\{Y_t\}$ for all $t=1,\ldots,T$. 

\subsection{Power-law coherency estimators}

Power-law coherency in time domain can be seen as a convergence to zero squared correlation for high scales $s$ as represented by Eqs. \ref{eq_PLC_time}-\ref{eq_PLC}. Following the ideas of Refs. \cite{Zebende2011,Kristoufek2014b}, the scale-specific covariance $\sigma_{xy}(s)$ and variances $\sigma_x^2(s)$ and $\sigma_y^2(s)$ can be directly substituted by either $F^2_{DCCA}(s)$ or $F^2_{DMCA}(\kappa)$ or $K_{xy}(\tau)$ based on DCCA, DMCA and HXA, respectively, into Eq. \ref{eq_PLC_time} so that we obtain a scale-specific squared correlation for each method as
\begin{gather}
\rho_{DCCA}^2(s)=\frac{|F_{DCCA}^2(s)|^2}{F_{DFA,x}^2(s)F_{DFA,y}^2(s)}, \nonumber \\ 
\rho_{DMCA}^2(\kappa)=\frac{|F_{DMCA}^2(\kappa)|^2}{F_{DMA,x}^2(\kappa)F_{DMA,y}^2(\kappa)}, \nonumber \\
\rho_{HXA}^2(\tau)=\frac{|K_{xy}(\tau)|^2}{K_{x}(\tau)K_{y}(\tau)}.
\label{eq:PLC1}
\end{gather}
Substituting Eqs. \ref{eq:DCCA3}, \ref{eq:CC-DMA_scaling} and \ref{eq:MFHXAeq2} into the above squared correlations, we get the following scaling laws:
\begin{gather}
\rho_{DCCA}^2(s) \propto \frac{s^{4H_{xy}}}{s^{2H_x}s^{2H_y}}=s^{4H_{xy}-2H_x-2H_y}=s^{4H_{\rho}} \nonumber \\ 
\rho_{DMCA}^2(\kappa) \propto \frac{\kappa^{4H_{xy}}}{\kappa^{2H_x}\kappa^{2H_y}}=\kappa^{4H_{xy}-2H_x-2H_y}=\kappa^{4H_{\rho}} \nonumber \\
\rho_{HXA}^2(\tau) \propto \frac{\tau^{4H_{xy}}}{\tau^{2H_x}\tau^{2H_y}}=\tau^{4H_{xy}-2H_x-2H_y}=\tau^{4H_{\rho}}
\label{eq:PLC2}
\end{gather}

The estimate of the power-law coherency parameter $H_{\rho}$ can be obtained via a log-log regression. In Eq. \ref{eq:PLC1}, the squared covariances in the numerators ensure a stable power-law scaling. For DCCA and DMCA, the scaling holds for high scales $s$ and $\kappa$, respectively, and for low differencing levels $\tau$ for HXA. The transition from power-law correlations and cross-correlations to power-law coherency is thus very straightforward. Note that $\rho_{DCCA}^2(s)$ and $\rho_{DMCA}^2(\kappa)$ come directly as squared scale-dependent correlation coefficients developed in the literature earlier \cite{Kristoufek2014a,Kristoufek2014b}. 

\section{Finite sample properties}

\subsection{Simulations setting}

We are interested in performance of the proposed estimators of power-law coherency. For this purpose, we utilize the mixed-correlated ARFIMA processes framework \cite{Kristoufek2013} which allows for controlling parameter $H_{\rho}$ via controlling $H_x$, $H_y$ and $H_{xy}$. Defining $a_n(d_i)=\frac{\Gamma(n+d_i)}{\Gamma(n+1)\Gamma(d_i)}$ for specific $d_i=H_i-0.5$, the mixed-correlated ARFIMA processes are defined as
\begin{gather}
x_t=\sum_{n=0}^{+\infty}{a_n(d_1)\varepsilon_{1,t-n}}+\sum_{n=0}^{+\infty}{a_n(d_2)\varepsilon_{2,t-n}} \nonumber \\
y_t=\sum_{n=0}^{+\infty}{a_n(d_3)\varepsilon_{3,t-n}}+\sum_{n=0}^{+\infty}{a_n(d_4)\varepsilon_{4,t-n}} \nonumber\\
\langle \varepsilon_{i,t} \rangle = 0\text{ for }i=1,2,3,4 \nonumber\\
\langle \varepsilon_{i,t}^2 \rangle = \sigma_{\varepsilon_i}^2\text{ for }i=1,2,3,4 \nonumber\\
 \langle \varepsilon_{i,t}\varepsilon_{j,t-n} \rangle = 0\text{ for }n \ne 0\text{ and }i,j=1,2,3,4 \nonumber\\
\langle \varepsilon_{i,t}\varepsilon_{j,t} \rangle = \sigma_{ij}\text{ for }i,j=1,2,3,4\text{ and }i\ne j.
\label{eq:ARFIMA_LC}
\end{gather}
The power-law coherency is obtained when $\{\varepsilon_2\}$ and $\{\varepsilon_3\}$ are correlated, and the other pairs are uncorrelated. In addition, we need $d_1>d_2$ and $d_4>d_3$ so that we get $H_{xy}<\frac{H_x+H_y}{2}$ \cite{Kristoufek2014,Kristoufek2015a}. Specifically, we have $H_x=d_1+0.5$, $H_{y}=d_4+0.5$ and $H_{xy}=0.5+\frac{1}{2}(d_2+d_3)$, or in other words $H_x=H_1$, $H_{y}=H_4$ and $H_{xy}=\frac{1}{2}(H_2+H_3)$. In the simulations, we fix $d_1=d_4=0.4$ and $d_2=d_3=0.2$ and the theoretical Hurst exponents are thus equal to $H_x=H_y=0.9$ and $H_{xy}=0.7$. We study three different time series lengths -- $T=500,1000,5000$ -- and we are interested in the effect of the strength of the correlation between the error terms $\{\varepsilon_2\}$ and $\{\varepsilon_3\}$. To do so, we check the finite sample properties for the correlation levels 0.1, 0.5 and 0.9. For each setting, we perform 1,000 repetitions.

As the estimation of the power-law coherency parameter $H_{\rho}$ is based on the log-log regression, we need to specify a range of scales included in the regression. For DCCA, we set $s_{max}=T/5$, which is standardly done in the literature, and we manipulate $s_{min}=10,20,50$ for $T=500$ and $s_{min}=10,50,100$ for the other two cases, $T=1000,5000$. For DMCA, the minimum moving average window size is set to $\kappa_{min}=3$ and the maximum one is investigated for three levels $\kappa_{max}=21,51,101$. For HXA, we are interested in its performance using various maximum scales $\tau_{max}$. In accordance with the literature \cite{DiMatteo2003,DiMatteo2005,DiMatteo2007,Kristoufek2011}, we use $\tau_{max}=20$ as a starting point and to obtain more stable estimates, we apply the jackknife procedure which estimates the power-law coherency parameter as an average of these estimates based on $\tau_{min}=1$ and $\tau^{\ast}_{max}=5,\ldots,\tau_{max}$. In addition, we check the maximum scales of 50 and 100 as well.

\subsection{Results}

We present results of the simulations based on the setting described above looking at the bias, variance and mean squared error (MSE, the sum of squared bias and variance) of the estimators.

The DCCA-based method does not attain desirable properties as shown in \autoref{tab:DCCA_rho}. For low correlation between error-terms $\varepsilon_2$ and $\varepsilon_3$, the bias reaches values above 0.4 and the situation does not improve much for the higher correlations mainly due to very high variance of the estimator. Even for the best case, which is the shortest series with $s_{min}=10$ and correlation between innovations equal to 0.9, we have a relatively low bias of 0.03 but a standard deviation of the estimator is still equal to approximately 0.2. Nonetheless, the bias and variance decrease considerably with the strength of correlation between innovations. Both bias and variance increase with the increasing minimum scale $s_{min}$. We cannot generally say that the bias and variance decrease with the time series length as these results vary for different combinations of $s_{min}$ and correlation between $\varepsilon_2$ and $\varepsilon_3$. All in all, the DCCA-based estimator of $H_{\rho}$ does not give satisfying results.

Performance of the HXA-based estimator of power-law coherency is summarized in \autoref{tab:HXA_rho}. Even though the estimator practically collapses for the low levels of correlation between $\varepsilon_2$ and $\varepsilon_3$ in a similar manner as the DCCA-based estimator does, the situation improves rapidly for the higher correlation levels. Very importantly, the bias of the estimator decreases with the time series length, which is a huge improvement over the DCCA-based estimator. The variance of the estimator decreases with the time series length as well, supporting the previous finding. Increasing the maximum scale increases the bias but decreases the variance of the estimator in most cases. However, the decrease in variance does not offset the increasing bias as the mean squared error increases with the maximum scale (apart from the least correlated cases).

For the DMCA-based method, we observe that the bias and variance decrease with the time series length but increase with $\kappa_{max}$. The bias also decreases with increasing correlation between error-terms $\varepsilon_2$ and $\varepsilon_3$ while the variance increases slightly. The DMCA approach strongly outperforms both the DCCA and HXA methods in this aspect (\autoref{tab:DMCA_rho}). Even though the increasing scale $\kappa_{max}$ has non-monotonous effects on bias (compare e.g. $\rho=0.9$ with $\kappa_{max}=21$, $\kappa_{max}=51$ and $\kappa_{max}=101$), it is evident that the mean squared error is the lowest for the lowest maximum scale (apart from the very weakly correlated series with $\rho=0.1$).

\begin{figure}[!htbp]
\begin{center}
\begin{tabular}{cc}
\includegraphics[width=80mm]{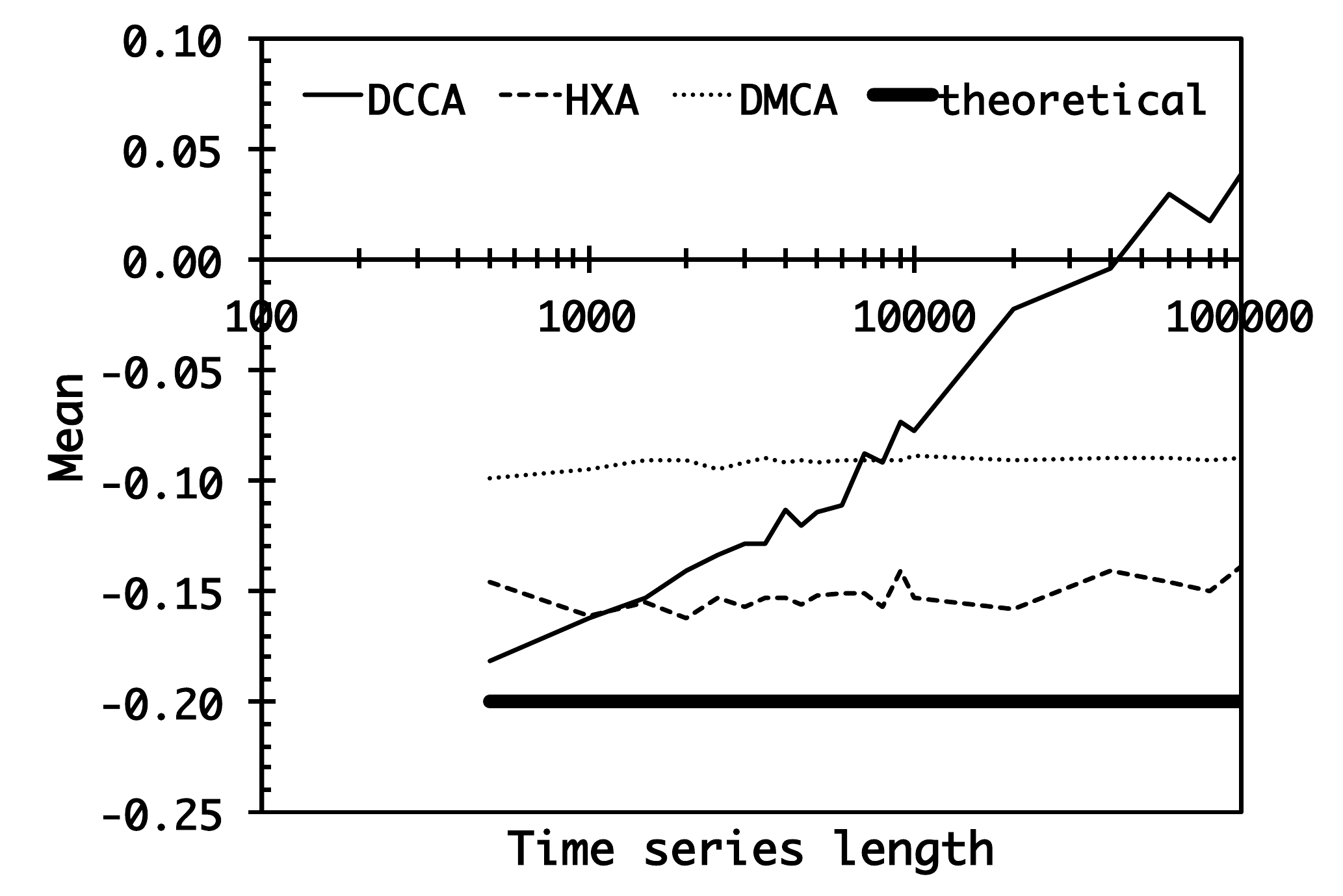}&\includegraphics[width=80mm]{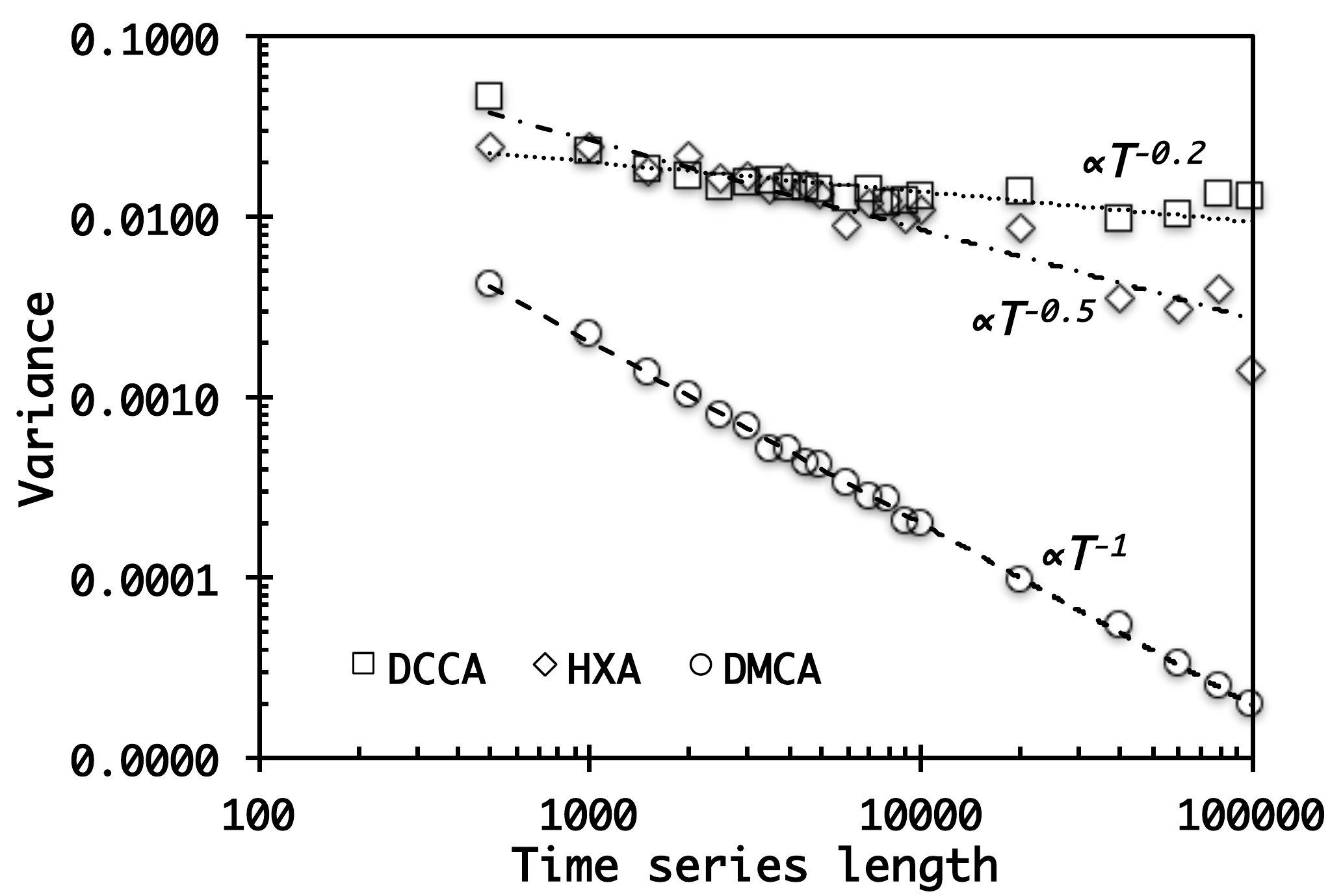}\\
\end{tabular}
\begin{tabular}{c}
\includegraphics[width=80mm]{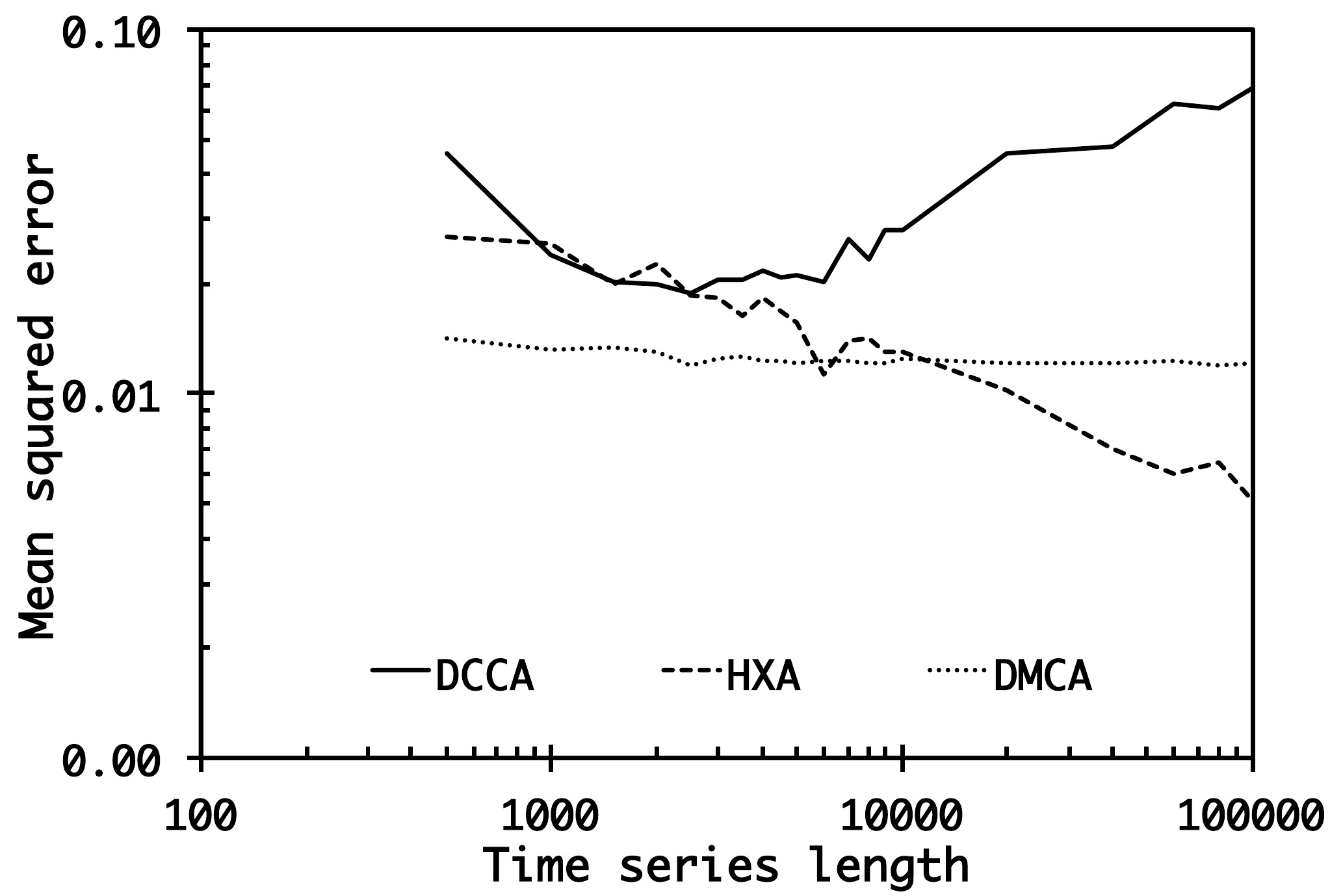}
\end{tabular}
\end{center}\vspace{-0.5cm}
\caption[Comparison of DCCA, HXA and DMCA power-law coherency estimators]{\textbf{Comparison of DCCA, HXA and DMCA power-law coherency estimators.} \footnotesize{Values are based on 1,000 simulations of mixed-correlated ARFIMA(0,$d$,0) processes with $d_1=d_4=0.4$, $d_2=d_3=0.2$ and perfectly correlated error-terms $\varepsilon_2$ and $\varepsilon_3$. Time series length ($x$-axis) varies between $500$ and $100 000$. The thick line represents the true value of $H_{\rho}=-0.2$. The DCCA-based method is the least biased for short time series but its bias increases logarithmically with time series length. The other two methods show a stable bias practically independent of the time series length. From the variance perspective, the DMCA-based method dominates the other two methods with a rapidly decaying variance at a rate of -1. When bias and variance are combined into the mean squared error, the DCCA-based method turns out to the worst of the three. The DMCA method outperforms the HXA method up to $T=5000$ where the latter method overtakes.}\label{fig:time_3}
}
%\begin{source}author's computations.\end{source}
\end{figure}

To provide an additional comparison between the methods, we again use the mixed-correlated ARFIMA processes with $d_1=d_4=0.4$, $d_2=d_3=0.2$ but we set for perfectly correlated $\varepsilon_2$ and $\varepsilon_3$ and uncorrelated other pairs. This gives us the theoretical power-law coherency of $H_{\rho}=-0.2$. For the specific methods, we use the parameter setting that proved the best in the previous simulations -- $s_{min}=10$, $\kappa_{max}=21$ and $\tau_{max}=20$. In addition, the time series length span is expanded up to $T=10^5$ (i.e. the span is between $500$ and $100 000$ observations) for better discussion about convergence of the methods towards the true value or the decay of variance. \autoref{fig:time_3} summarizes the results of simulations, namely bias, variance and mean squared error dependence on time series length. 

The HXA and DMCA methods show a stable upward bias of approximately 0.05 and 0.1, respectively, regardless the time series length. The bias increases with the time series length for DCCA. As the time series length axis is shown in a logarithmic scale and the mean value of the estimator in a linear one, the bias of the DCCA-based method increases approximately logarithmically. With respect to variance, the DMCA power-law coherency estimator clearly outperforms the other two estimators. In the right panel of Figure \ref{fig:time_3}, we observe a log-log plot of variance with respect to the time series length. A clear power-law scaling emerges for the DMCA method with a slope of -1, i.e. the variance of the estimator decays as $\frac{1}{T}$, which is a target rate for consistent estimators. The HXA and DCCA estimators attain very similar levels of variance for $T<10000$. For longer time series, the HXA method dominates the DCCA method. This is reflected in different decay rates for each method. For the DCCA-based power-law coherency estimator, we find a power-law scaling with an exponent of 0.2, i.e its variance decays very slowly at a rate of $\frac{1}{T^{0.2}}$. The HXA-based estimator performs in between the other two estimators with a rate of 0.5, i.e. its variance decays with a rate of $\frac{1}{\sqrt{T}}$.

Putting bias and variance together, the mean squared error shows an interesting behavior. For short series with $T<5000$, the DMCA-based method clearly dominates. However, its MSE is quite stable with changing time series length showing only a very mild decrease. This only reflects the fact that its bias is rather stable, and its variance is very low even for short time series lengths. The DCCA-based method is the worst one of the three for practically all examined time series lengths. After a decreasing trend for very short series with $T<1000$, MSE plateaus between $1000 \le T \le 5000$ and increases for longer time series lengths, which is very undesirable. The HXA-based estimator shows similar levels of MSE as the DCCA-based one for short series with $T<3000$. Contrary to the DCCA estimator, it follows the decreasing trend of MSE even for higher $T$ and it even outperforms the DMCA-based method for $T>10000$.

\section{Conclusion}

We have focused on power-law coherency as an alternative approach towards studying power-law cross-correlations between simultaneously recorded time series. To be able to study empirical data, we have introduced three estimators of the power-law coherency parameter $H_{\rho}$ based on popular techniques utilized for studying power-law cross-correlations -- detrended cross-correlation analysis, detrending moving-average cross-correlation analysis and height cross-correlation analysis. In the finite sample properties study focusing on the bias, variance and mean squared error of the estimators, we have uncovered several interesting findings. First, the DCCA-based estimator performs the worst of the three studied methods. Specifically, its bias increases with the time series length and this is not offset enough by its slowly decaying variance. Second, the other two methods (based on HXA and DMCA) have a stable bias which does not change with the time series length. Third, the variance decay differs strongly for the examined methods. The DMCA-method strongly outperforms the other two as it starts at lower variance levels and decays with the most rapid rate. The other two methods have similar levels of variance but the HXA method has a faster rate of variance decay. Overall, the DMCA-based method is a safe choice among the three. The HXA method is reasonable for long time series with at least $10^4$ observations, which can be easily attainable in some disciplines but problematic in others. The DCCA-based method does not provide favorable properties which even deteriorate with an increasing time series length. %These results might serve as a basis for further research in the domain of power-law coherency.

\section*{Acknowledgements}
%The author would like to thank the anonymous referees for valuable comments and suggestions which helped to improve the paper significantly. 
%The research leading to these results has received funding from the European Union's Seventh Framework Programme (FP7/2007-2013) under grant agreement No. FP7-SSH-612955 (FinMaP). 
Support from the Czech Science Foundation under project No. 14-11402P is gratefully acknowledged.

\newpage

\section*{References}
%\bibliography{Bibliography}
%\bibliographystyle{unsrt}

\newpage

\begin{landscape}

\begin{table}[!htbp]
\caption{\textbf{Finite sample properties of the DCCA estimator.} DCCA estimator of $H_{\rho}$ for Mixed-correlated ARFIMA processes with $d_1=d_4=0.4$, $d_2=d_3=0.2$ and varying correlation between $\varepsilon_2$ and $\varepsilon_3$.\label{tab:DCCA_rho}}
\centering
%\tiny
\begin{tabular}{|c|c|ccc|ccc|ccc|}
\hline \hline
&&&$\rho=0.1$&&&$\rho=0.5$&&&$\rho=0.9$&\\
\hline
&&bias&SD&MSE&bias&SD&MSE&bias&SD&MSE\\
\hline \hline
&$s_{min}=10$&0.4321&0.2738&0.2617&0.1628&0.2255&0.0774&0.0333&0.1953&0.0393\\
$T=500$&$s_{min}=20$&0.4427&0.3888&0.3471&0.2404&0.3768&0.1998&0.0345&0.2879&0.0841\\
&$s_{min}=50$&0.4418&0.9231&1.0472&0.3526&0.9021&0.9312&0.1317&0.7669&0.6054\\
\hline
&$s_{min}=10$&0.4346&0.1987&0.2283&0.1660&0.1590&0.0528&0.0506&0.1461&0.0239\\
$T=1000$&$s_{min}=50$&0.4686&0.4156&0.3924&0.3141&0.3975&0.2567&0.1174&0.3487&0.1354\\
&$s_{min}=100$&0.4669&0.7759&0.8201&0.3886&0.8120&0.8103&0.2226&0.7477&0.6087\\
\hline
&$s_{min}=10$&0.4217&0.1277&0.1942&0.2194&0.1245&0.0636&0.1123&0.1120&0.0252\\
$T=5000$&$s_{min}=50$&0.4530&0.1794&0.2374&0.2873&0.1892&0.1183&0.1474&0.1658&0.0492\\
&$s_{min}=100$&0.4634&0.2260&0.2658&0.3405&0.2450&0.1760&0.1943&0.2269&0.0892\\
\hline \hline
\end{tabular}
\end{table}

\begin{table}[!htbp]
\caption{\textbf{Finite sample properties of the HXA estimator} HXA estimator of $H_{\rho}$ for Mixed-correlated ARFIMA processes with $d_1=d_4=0.4$, $d_2=d_3=0.2$ and varying correlation between $\varepsilon_2$ and $\varepsilon_3$.\label{tab:HXA_rho}}
\centering
%\tiny
\begin{tabular}{|c|c|ccc|ccc|ccc|}
\hline \hline
&&&$\rho=0.1$&&&$\rho=0.5$&&&$\rho=0.9$&\\
\hline
&&bias&SD&MSE&bias&SD&MSE&bias&SD&MSE\\
\hline \hline
&$\tau_{max}=20$&0.3797&0.2153&0.1905&0.1446&0.2178&0.0683&0.0556&0.1675&0.0312\\
$T=500$&$\tau_{max}=50$&0.3873&0.1605&0.1758&0.2095&0.1746&0.0744&0.1057&0.1409&0.0310\\
&$\tau_{max}=100$&0.4016&0.1324&0.1788&0.2614&0.1442&0.0892&0.1664&0.1222&0.0426\\
\hline
&$\tau_{max}=20$&0.3636&0.2209&0.1810&0.0858&0.1977&0.0465&0.0439&0.1586&0.0271\\
$T=1000$&$\tau_{max}=50$&0.3715&0.1681&0.1663&0.1563&0.1667&0.0522&0.0730&0.1468&0.0269\\
&$\tau_{max}=100$&0.3796&0.1402&0.1638&0.2111&0.1502&0.0671&0.1211&0.1344&0.0327\\
\hline
&$\tau_{max}=20$&0.2956&0.2512&0.1505&0.0424&0.1739&0.0320&0.0456&0.1216&0.0169\\
$T=5000$&$\tau_{max}=50$&0.3160&0.1924&0.1369&0.0818&0.1572&0.0314&0.0444&0.1278&0.0183\\
&$\tau_{max}=100$&0.3317&0.1550&0.1341&0.1288&0.1477&0.0384&0.0623&0.1322&0.0214\\
\hline \hline
\end{tabular}
\end{table}

\begin{table}[!htbp]
\caption{\textbf{Finite sample properties of the DMCA estimator.} DMCA estimator of $H_{\rho}$ for Mixed-correlated ARFIMA processes with $d_1=d_4=0.4$, $d_2=d_3=0.2$ and varying correlation between $\varepsilon_2$ and $\varepsilon_3$.\label{tab:DMCA_rho}}
\centering
%\tiny
\begin{tabular}{|c|c|ccc|ccc|ccc|}
\hline \hline
&&&$\rho=0.1$&&&$\rho=0.5$&&&$\rho=0.9$&\\
\hline
&&bias&SD&MSE&bias&SD&MSE&bias&SD&MSE\\
\hline \hline
&$\kappa_{max}=21$&0.2652&0.4064&0.2355&0.0723&0.1976&0.0443&0.0973&0.0834&0.0164\\
$T=500$&$\kappa_{max}=51$&0.3622&0.3062&0.2249&0.0360&0.2404&0.0591&0.0442&0.1759&0.0329\\
&$\kappa_{max}=101$&0.4117&0.2580&0.2361&0.0920&0.2202&0.0570&0.0293&0.1998&0.0408\\
\hline
&$\kappa_{max}=21$&0.1942&0.4033&0.2004&0.0891&0.1229&0.0230&0.1062&0.0540&0.0142\\
$T=1000$&$\kappa_{max}=51$&0.2893&0.2994&0.1733&0.0355&0.2106&0.0456&0.0721&0.1022&0.0156\\
&$\kappa_{max}=101$&0.3453&0.2706&0.1925&0.0440&0.2148&0.0481&0.0306&0.1749&0.0315\\
\hline
&$\kappa_{max}=21$&0.0640&0.2816&0.0834&0.1056&0.0475&0.0134&0.1096&0.0223&0.0125\\
$T=5000$&$\kappa_{max}=51$&0.1097&0.2658&0.0826&0.0821&0.0799&0.0131&0.0917&0.0369&0.0098\\
&$\kappa_{max}=101$&0.1806&0.2358&0.0882&0.0494&0.1402&0.0221&0.0732&0.0644&0.0095\\
\hline \hline
\end{tabular}
\end{table}

\end{landscape}

\end{document}